\documentclass[prd,a4paper,twocolumn,superscriptaddress]{revtex4}
\usepackage{graphicx}
 
\newcommand{\be}{\begin{equation}}
\newcommand{\ee}{\end{equation}}
\newcommand\lsim{\mathrel{\rlap{\lower4pt\hbox{\hskip1pt$\sim$}}
    \raise1pt\hbox{$<$}}}
\newcommand\gsim{\mathrel{\rlap{\lower4pt\hbox{\hskip1pt$\sim$}}
    \raise1pt\hbox{$>$}}}
\newcommand\esim{\mathrel{\rlap{\raise2pt\hbox{\hskip0pt$\sim$}}
    \lower1pt\hbox{$-$}}}

\begin{document}

\title{Scaling defect decay and the reionization history of the Universe}

\author{P.P. Avelino}
\email[Electronic address: ]{ppavelin@fc.up.pt}
\affiliation{Centro de F\'{\i}sica do Porto, Rua do Campo Alegre 687,
4169-007, Porto, Portugal}
\affiliation{Departamento de F\'{\i}sica da Faculdade de
Ci\^encias da Universidade do Porto, Rua do Campo Alegre 687,
4169-007, Porto, Portugal}
\author{D. Barbosa}
\email[Electronic address: ]{barbosa@supernova.ist.utl.pt}
\affiliation{CENTRA, Instituto Superior T\'ecnico, 
Av. Rovisco Pais, 1049-001 Lisboa Codex, Portugal}
 
\date{2 June 2004} 
\begin{abstract} 

We consider a model for the reionization history of the Universe in
which a significant fraction of the observed optical depth is a result
of direct reionization by the decay products of a scaling cosmic defect
network. We show that such network can make a significant contribution to
the reionization history of the Universe even if its energy density is
very small (the defect energy density has to be greater than about 
$10^{-11}$ of the background density). We compute the Cosmic Microwave
Background temperature, polarization and temperature-polarization 
cross power spectrum and show that
a contribution to the observed optical depth due to the decay
products of a scaling defect network may help to reconcile a high
optical depth with a low redshift of complete reionization suggested by quasar
data. However, if the energy density of defects is approximately a constant
fraction of the background density then these models do not explain the
large scale bump in the temperature-polarization cross power spectrum 
observed by WMAP.

\end{abstract} 
\preprint{tba} 
\maketitle 
 
\section{Introduction} 

Two important types of observations have recently begun to probe the 
reionization history of the Universe. On one hand the spectra of high 
redshift quasars at $z \sim 6$ revealed the presence of 
Gunn-Peterson troughs \cite{Gunn} suggesting a very late reionization 
\cite {Becker,Fan,White,Gnedin}, a fact corroborated by the possible detection 
of first reionization sources around this redshift \cite{Stiavelli}. On 
the other hand the measurement by the 
WMAP satellite \cite{Kogut,Spergel} of a large correlation between the 
temperature and 
E-type polarization at large angular scales seems to indicate a high optical 
depth which was interpreted as evidence for early reionization 
($z_{\rm reion}=11-30$). These two apparently contradictory results suggest 
that the reionization history of the Universe may be more complex that 
originally thought. In ref. \cite{Gnedin} 
it was shown that numerical simulations in the context of a 
``classical model'' for cosmological reionization lead to a consistent 
picture between SLOAN data and WMAP provided there is a slow partial 
reionization from early epochs reaching total ionization by
$z \sim 6$. A number of astrophysical and particle physics models 
that may help to reconcile CMB and quasar results have recently been 
suggested. For example, in ref. \cite{Centwice, Cen} 
a double-reionization model in which a first stage of reionization by metal 
free stars at high redshift is followed by a second one at lower redshifts 
was studied in detail, while other models with more than one relevant source 
of reionization have also been investigated \cite{Hui,Ciardi,Madau}. An 
alternative 
explanation may 
lie on yet unknown fundamental physics which might be probed by reionization.
For example, it was suggested \cite{Avelino, Chen} that a slower reionization 
with a high optical depth might be achieved in the context of 
non-gaussian models either in the context of inflation \cite{Peebles} of 
topological 
defects \cite{Avelino1}. Another possibility that has 
recently received much attention is the decay of dark matter particles which 
may release the energy necessary to reionize the intergalactic medium 
\cite{Hansen,Kasuya,Pierpaoli,Kamionkowski}.

In this article we consider yet another possibility for explaining both 
WMAP and quasar results in which a significant fraction of the observed 
optical depth is a result of direct reionization by the decay products 
of a scaling network of topological defects. We will make our model as general 
as possible looking at specific signatures of a scaling solution for 
defect evolution. Scaling is a generic feature of defect models in which 
the statistical properties of the defect network remain self-similar 
relative to the Hubble radius. In the most interesting models it implies that 
the defect energy density is roughly proportional to the background 
density in the radiation and matter eras. This means that the scaling 
defect network will loose a roughly constant fraction of its energy each 
Hubble time. Part of the energy lost by the network may provide a useful 
contribution to the reionization history of the Universe leaving specific 
signatures in the CMB which we investigate in this paper.

The article is organized as follows. In section II we describe our model and 
discuss the appropriate modifications to the standard reionization history. 
In section III we calculate the evolution of the ionization fraction with 
redshift, the optical depth, the temperature, E-type polarization and 
temperature-polarization cross power spectra (TE) for our model comparing 
with WMAP observations and discussing the results in detail. We summarise 
our results in section IV.
 
\section{The model} 
 
We consider a scaling network \cite{Book} for which the energy density, 
$\rho_D$, is a constant fraction of the background density, $\rho_b$. 
Given that in the matter era $\rho_b \propto t^{-2}$ we have 
\begin{equation} 
\frac{d \rho_D}{dt}= -2 \frac{\rho_D}{t}\, ,
\end{equation} 
where $t$ is the physical time.
We shall investigate a non-standard scenario in which the decay products 
of a scaling defect network may act as an additional source of 
ionization. The nature of decay 
particles produced by a defect network is model dependent and 
may include a large variety of more or less exotic particles, some 
of which can in principle make a significant contribution 
to the reionization history of the Universe. We will assume that a 
fraction $\beta$ of the energy lost by the network will affect the 
reionization history 
either through direct ionization (a fraction $\epsilon$) or by heating 
the intergalactic medium (a fraction $1-\epsilon$). The parameter $\beta$ 
must be smaller than unity and may be even be much smaller than that 
depending on the relative importance of the decay mechanisms available 
to the defect network. For example, if gravitational radiation is the 
dominant decay channel (as appears to be the case for Abelian Higgs 
networks \cite{Moore} - see however ref. \cite{Graham}) $\beta$ 
will necessarily be small.

In refs. \cite{Pierpaoli,Kamionkowski} and following 
the work of ref. \cite{Shull}  $\epsilon=1/3$ was taken as a reasonable 
approximation to the fraction of the energy that goes into 
ionization for the case where electrons heat the IGM when the 
gas is mostly neutral. We note that we want our model to remain as general 
as possible in order to look at specific reionization signatures of a 
scaling solution for defect evolution. However, our 
results  are weakly dependent on the specific value of $\epsilon$ and so we 
will make the best motivated choice ($\epsilon=1/3$).

Here we will also make the simplifying assumption that the fraction of the 
energy that goes into excitations is negligible and neglect helium 
ionization. Again this does not greatly affect our results and so we choose 
to avoid unnecessary complexity.

\begin{figure} 
\includegraphics[width=3.5in,keepaspectratio]{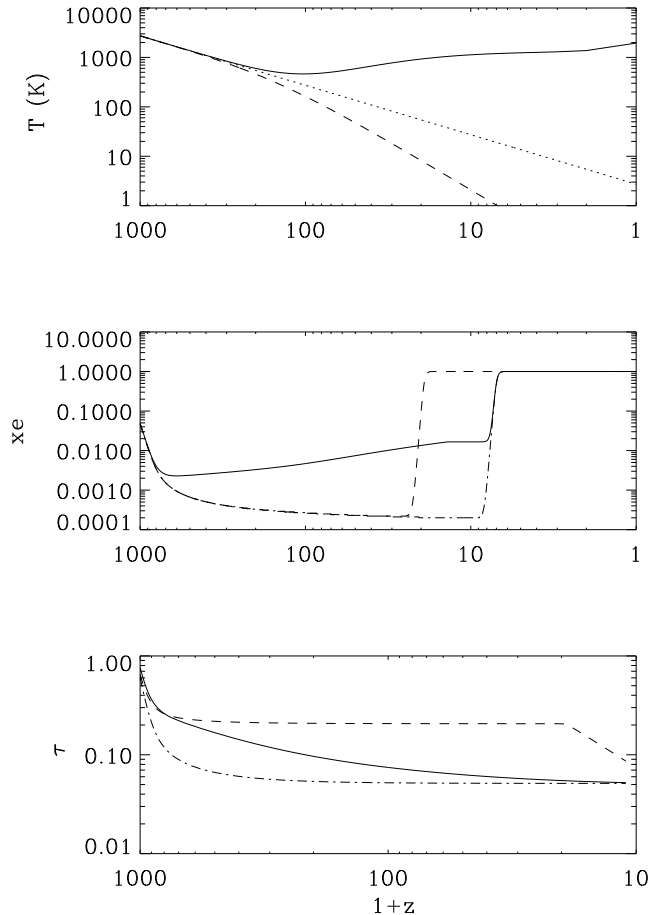} 
\caption{\label{fig1} The IGM temperature, ionization 
fraction and optical depth for models with fast reionization at 
$z \sim 17$ (model I - dashed line) and $z \sim 6$ (model II - dot-dashed 
line) induced by standard sources and the scaling defect model with parameters 
$\epsilon=1/3$ and $\Omega_D^0=2 \beta^{-1} \times 10^{-11}$ with rapid 
reionization at $z \sim 6$ by standard sources (model III - solid line). In 
the first plot the dotted line represents the evolution of the CMB 
temperature.}
\end{figure} 

The defect network will provide an additional ionization source which 
should be added to the standard ionization equation 
\begin{equation} 
\label{two}
\frac{d n_e}{dt}=  \frac{2 \epsilon \beta \rho_D}{I_H t}\, ,
\end{equation} 
where $n_e$ is the number density of free electrons and $I_H=13.6 {\rm eV}$. 
Part of this energy will contribute to 
heat the intergalactic medium and to this end we added an additional 
term to the equation which describes the evolution of the gas temperature, $T$
\begin{equation}
\label{three} 
\frac{d T}{dt}=  \frac{4 \beta \rho_D}{3 n k_B t}
\left(1-\epsilon-\epsilon\frac{3 k_B T}{2I_H}\right)\, ,
\end{equation} 
where $n$ is the particle number density for the gas and $k_B$ is the 
Boltzmann constant. We modified the RECFAST code \cite{Seager} by adding the 
terms in equations (\ref{two}-{three}) and obtained the temperature, E-type 
polarization and cross power spectra using CMBFAST \cite{Lewis}.

\section{Results and discussion} 

\begin{figure} 
\includegraphics[width=3.5in,keepaspectratio]{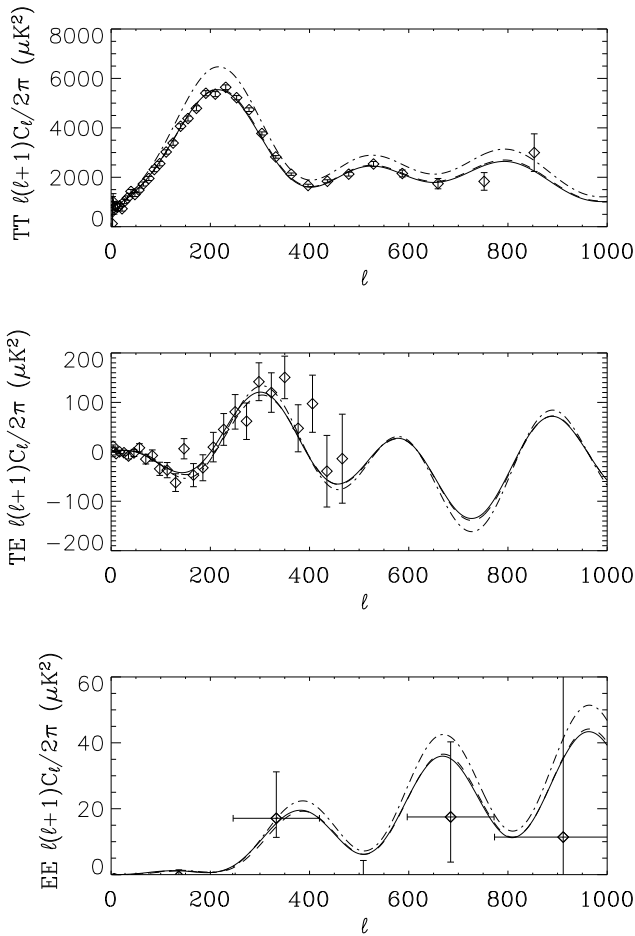} 
\caption{\label{fig2} The CMB temperature (TT), polarization (EE) and 
cross power spectra (TE) for models I (dashed line), II (dot-dashed line) 
and III (solid line).
The data points with error bars for the TT and TE spectra are the binned 
data given by the WMAP team while the error bars for 
the EE power spectrum are the binned data given by the DASI team. We see 
that the models I and III lead to very similar power spectra.} 
\end{figure} 

\begin{figure} 
\includegraphics[width=3.5in,keepaspectratio]{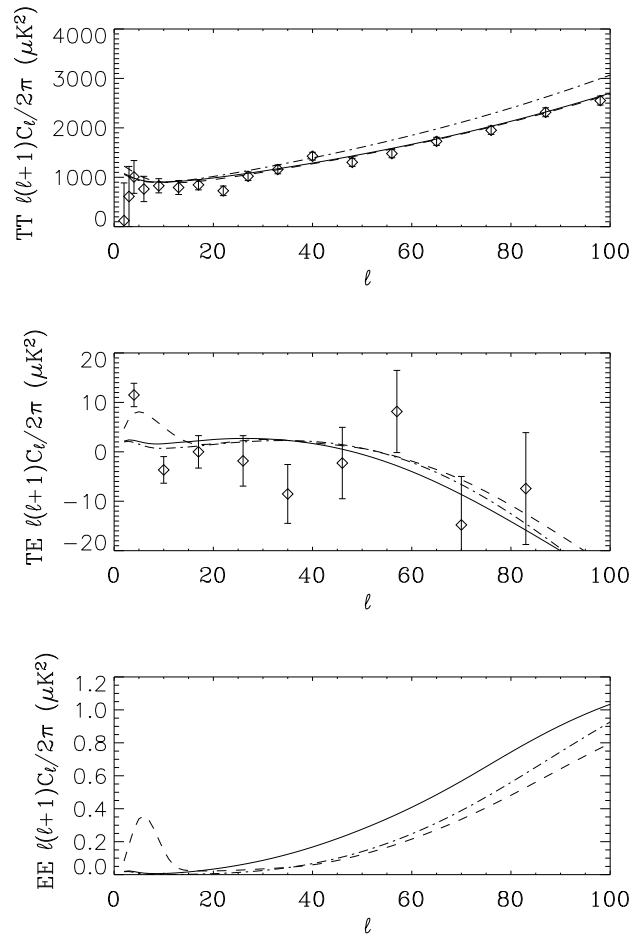} 
\caption{\label{fig3} Same as Fig. \ref{fig2} but for $\ell \le 100$ 
We see that direct reionization due to the decay products of a scaling 
defect network is not able to explain the low $\ell$ bump in TE cross 
power spectrum.} 
\end{figure} 

We now discuss in more detail how the decay products of a 
scaling cosmic defect network may affect the reionization history of the 
Universe. Throughout, the paper we shall adopt a cosmological model 
motivated by the WMAP results (Bennett et 
al.~2003) with initial gaussian fluctuations and a matter density 
$\Omega_{{\rm m}}^0=0.29$, dark energy density $\Omega_\Lambda^0=0.71$,
baryon density $\Omega_{{\rm B}}^0=0.047$, Hubble parameter $h=0.72$, 
normalization $\sigma_8=0.9$, and perturbation spectral index $n_{{\rm 
s}}=0.99$. Although there may be an important contribution to the 
reionization history of the Universe coming non-gaussian perturbations 
\cite{Avelino, Chen}
we shall assume throughout this paper that the energy density of the 
defect network is very small so that it does not contribute significantly 
to structure formation. We calculate the IGM temperature, 
ionization fraction and optical depth for a number of models considering 
in particular fast reionization at $z \sim 17$ (model I - dashed line) and 
$z \sim 6$ (model II - dot-dashed line) 
induced by standard sources and a 
scaling defect model with parameters $\epsilon=1/3$ and 
$\Omega_D^0=2 \beta^{-1} \times 10^{-11}$ with standard reionization 
induced by standard sources occurring at $z \sim 6$ (model III -solid line). 
Note that in the particular case of cosmic strings model III would require a 
string mass per unit length $G\mu \sim \beta^{-1} 10^{-13}$ making it 
possible (if $\beta$ is large enough) for cosmic strings to influence the 
reionization history of the Universe even if they play a negligible role 
in structure formation. 

In Fig.\ref{fig1} we plot the evolution of the IGM 
temperature, ionization fraction and optical depth for models 
I, II and III. The evolution 
of the IGM temperature is not to be trusted at small $z$ 
since the evolution of $x_e$ due to fast reionization by standard sources 
was put in by hand and no specific model for the evolution of the gas 
temperature was specified. We see that although models I and III 
have roughly the same optical depth up to $z \sim 800$ the reionization 
history is very different. In particular the residual ionization fraction 
is much larger in the case of model III and the evolution of the optical 
depth up to redshift $z$ is much smoother for that model which is related to 
the specific time dependence of the energy output associated with the 
scaling solution.

We note that these results are qualitatively in agreement with the results 
of \cite{Hansen,Kasuya,Pierpaoli,Kamionkowski} in which an 
important contribution to the reionization history of 
the Universe from decaying particles was studied. However, in our case 
$dn/dt \propto t^{-3}$ (instead of $dn/dt \propto t^{-2}$ caracteristic of 
decaying particles with a very long lifetime) which results in a slower 
evolution of $T$, $x_e$ and $\tau$ at low redshifts.

We compute the CMB temperature (TT), polarization (EE) and cross power 
spectra for models I (dashed line), II (dot-dashed line) and III (solid line). 
The results are 
displayed in Figs. \ref{fig2} (up to $\ell=100$) and \ref{fig3} 
(up to $\ell=1000$). The data points with error bars for the TT and TE 
spectra are the binned data given by the WMAP  team \cite{Kogut,Spergel} 
while the error bars for 
the EE power spectrum are the binned data given by the DASI 
team \cite{Kovac}. We see 
that models I and III lead to power spectra which are very similar showing 
that the decay products of a scaling defect network may help to reconcile a 
high optical depth with a low redshift of complete reionization suggested by 
quasar data. However we see in Fig. \ref{fig3} that model III does not explain 
the low $\ell$ bump in the TE cross power spectrum. We also find that 
even a value of $\Omega^0_D$ as low as $10^{-11}$ may lead to a  
significant contribution to the reionization history of the Universe depending 
on the value of $\beta$ which parametrizes the fraction of the energy lost 
by the scaling network which is useful to reionization.

\section{Conclusions}

In this article we studied the contribution to the reionization history 
of the Universe made by the decay products of a network of scaling 
cosmic defects with an energy density approximately a constant fraction 
of the background density. We calculated the TT, TE and EE power spectra 
showing that this model together with a standard epoch of fast reionization 
at $z \sim 6$ may explain the observed CMB results with exception of 
the low $\ell$ bump in the TE cross power spectrum.  In order to leave a 
significant imprint on the CMB of the defect network energy density 
has to be at least about $10^{-11}$ of the background density.

We note that not all scaling defect models have an energy density which 
scales roughly with the background density. In the particular case of 
a scaling network of domain walls $\rho_D/\rho_b \propto t$ in the 
radiation and matter eras. If domain walls are very light they may have a 
negligible direct impact on the CMB but their decay products could still 
be able to influence reionization. We verified that the time dependence of the 
energy input in this 
model is the same as that of decaying particles with very long lifetime 
previously studied in refs. \cite{Hansen,Kasuya,Pierpaoli,Kamionkowski}. 

We shall leave for future work a likelihood analysis for specific defect 
models taking into account the slow transition between different scaling 
regimes around radiation-matter equality as well as deviations from a 
scaling solution resulting from a late time acceleration of the Universe. 
We expect these features to produce only slight changes to the generic 
picture presented in this paper. This work seems favour a fast reionization 
of the Universe over a very slow one. However, we note that this conclusion 
is fully based on the low $\ell$ bump in the TE cross power spectra.

\begin{acknowledgments} 

We acknowledge Xuelei Chen for helpful comments on RECFAST.  
We acknowledge additional support from FCT under contracts 
CERN/POCTI/49507/2002 and POCTI/FNU/42263/2001. DB is also supported by FCT 
grant SFRH BPD/11640/2002.

\end{acknowledgments} 

\bibliography{reiond} 
\end{document}